\documentclass[letter,english]{article}
\usepackage{emulateapj,epsfig,onecolfloat}

\received{2002 May 24}
\begin{document}
\twocolumn
[
\title{Simulating the Sunyaev-Zel'dovich effect(s):
including radiative cooling and energy injection by galactic winds}
\author{Martin White${}^1$, Lars Hernquist${}^2$ and Volker Springel${}^3$}
\affil{${}^1$Departments of Astronomy and Physics, University of California,
Berkeley, CA 94720}
\affil{${}^2$Harvard-Smithsonian Center for Astrophysics, Cambridge, MA 02138}
\affil{${}^3$Max-Planck-Institut f\"{u}r Astrophysik,
Karl-Schwarzchild-Strasse 1, 85748 Garching, Germany}

\begin{abstract}
\noindent
\rightskip=0pt
We present results on the thermal and kinetic Sunyaev-Zel'dovich (SZ) effects
{}from a sequence of high resolution hydrodynamic simulations of structure
formation, including cooling, feedback and metal injection.
These simulations represent a self-consistent thermal model which incorporates
ideas from the `pre-heating' scenario while preserving good agreement with the
low density IGM at $z\sim 3$ probed by the Ly-$\alpha$ forest.
Four simulations were performed, at two different resolutions with and without
radiative effects and star formation.  The long-wavelength modes in each
simulation were the same, so that we can compare the results on an object by
object basis.
We demonstrate that our simulations are converged to the sub-arcminute level.
The effect of the additional physics is to suppress the mean Comptonization
parameter by $20\%$ and to suppress the angular power spectrum of fluctuations
by just under a factor of two in this model while leaving the source counts
and properties relatively unchanged.  We quantify how non-Gaussianity in the
SZ maps increases the sample variance over the standard result for Gaussian
fluctuations.
We identify a large scatter in the $Y$-M relation which will be important
in searches for clusters using the SZ effect(s).
\end{abstract}
  ]

\section{Introduction} \label{sec:intro}

On angular scales below $10'$, the microwave sky carries the imprint of
large-scale structure in the low-$z$ universe.  In particular, cosmic
microwave background (CMB) photons propagating through the universe
are inverse Compton or Doppler scattered by hot electrons along their
path, either in dense structures such as clusters of galaxies or more
generally in hot gas in the intergalactic medium.  Inverse Compton
scattering conserves the number of photons but preferentially
increases their energy, leading to a spectral distortion whose
amplitude is proportional to the product of the electron temperature
and density (or pressure).  Doppler scattering induces an intensity
fluctuation with the same spectral shape as the CMB itself.  These
effects were first described by Sunyaev \& Zel'dovich
(\cite{SZ72,SZ80}) and are known as the thermal and kinetic SZ
effects, respectively (for recent reviews see Rephaeli~\cite{Rep} and
Birkinshaw~\cite{Bir}).

The thermal effect is one of the primary sources of secondary
anisotropies in the CMB on small angular scales.  The change in the
(thermodynamic) temperature of the CMB resulting from scattering off
non-relativistic electrons is
\begin{eqnarray}
{\Delta T\over T} &=&
  \phantom{-2}y \left( x{{\rm e}^x+1\over {\rm e}^x-1}-4 \right) \\
  &\simeq& -2y\qquad \mbox{for }\ x\ll 1\, ,
\end{eqnarray}
where $x=h\nu/kT_{\rm CMB}\simeq \nu/56.85\;$GHz is the dimensionless
frequency, and the second expression is valid in the Rayleigh-Jeans
limit which we shall assume henceforth. The quantity $y$ is known as
the Comptonization parameter and is given by
\begin{equation}
y\equiv \sigma_T\int d\ell\ {n_e k(T_e-T_{\rm CMB})\over m_e c^2}\, ,
\label{eqn:ydef}
\end{equation}
where the integral is performed along the photon path.  Since $T_e\gg
T_{\rm CMB}$ the integrand is proportional to the integrated electron
pressure along the line of sight.

The kinetic SZ effect arises from the motion of ionized gas with
respect to the rest-frame of the CMB. The resulting temperature fluctuation
is ${\Delta T / T} = - b$, where
\begin{equation}
  b\equiv \sigma_T \int d\ell \ n_e {v_r\over c}
\end{equation}
measures the magnitude of the effect along the line of sight if $v_r$
is the radial peculiar velocity of the gas
(positive if the cluster is moving away from us, negative if it is moving
towards us).
The different dependence on frequency of the two effects can, in principle,
be used to disentangle them observationally, though the primary CMB
anisotropies provide ``noise'' for the kinetic SZ signal.

In this paper, we report results of a sequence of high resolution hydrodynamic
simulations designed to study the SZ signal on small angular scales.
The imprint of the SZ effects on the CMB has been studied by a number of
authors (Ostriker \& Vishniac \cite{OsVi}; Persi et~al.~\cite{PSCO};
da Silva et~al.~\cite{SBLT00,SBLT01}; Refregier et~al.~\cite{RKSP};
Seljak et~al.~\cite{SelBurPen}; Kay, Liddle \& Thomas~\cite{KayLidTho};
Refregier \& Teyssier~\cite{RefTey};
da Silva et al.~\cite{SKLTPB}; Zhang, Pen \& Wang~\cite{ZPW}).
Ours are the first such simulations to incorporate the effects of radiation
and star formation, including gas cooling and feedback from supernovae and
galactic winds.
As such they extend our earlier work on the subject, reported in
Springel, White \& Hernquist~(\cite{SprWhiHer}).

The outline of this paper is as follows.
In Section \ref{sec:method}, we briefly describe our simulations and the
techniques we use to compute SZ maps.
In Section \ref{sec:results}, we present our results for the power spectra
of thermal and kinetic SZ effects and discuss the effects of non-Gaussianity
on the sample variance.  We show SZ source counts as a function of source
strength and source size and how SZ source properties correlate with the
intrinsic properties of the cluster.
Finally, we summarize and discuss our results in Section \ref{sec:conclusions}.

\section{Method} \label{sec:method}

We chose as our cosmological model the currently favored $\Lambda$CDM
cosmology.  In particular, we adopted $\Omega_{\rm m}=0.3$,
$\Omega_\Lambda=0.7$, $H_0=100\,h\,{\rm km}\,{\rm s}^{-1}{\rm Mpc}^{-1}$
with $h=0.7$, $\Omega_{\rm B}=0.04$, $n=1$ and $\sigma_8=0.9$.
This model yields a reasonable fit to the current suite of cosmological
constraints and as such provides a good framework for making realistic
predictions.  The simulation domain was a $100\,h^{-1}$Mpc cube with periodic
boundary conditions.
We ran smoothed particle hydrodynamic (SPH) simulations at two different mass
resolutions corresponding to
$2\times 144^3$ and $2\times 216^3$ particles, with initial conditions
constructed such that the large-scale modes were equal for both
resolutions, allowing us to perform a resolution study on an object by
object basis without being affected by cosmic variance.  For each mass
resolution, we performed a simulation that computed only
adiabatic\footnote{In what follows, we will refer to the runs that ignored
radiative effects and star formation as `adiabatic' but it should be kept
in mind that these simulations did include shock heating and so are not
strictly speaking adiabatic.} gas physics and shock heating, and one that
also followed the radiative cooling and heating processes of the gas in
the presence of a UV background radiation field together with star formation
and feedback processes in the dense collapsed gas.
In Table~\ref{tab:sims}, we list some basic numerical parameters of the four
simulations that we carried out.

Radiative cooling and heating was modeled as in
Katz, Weinberg \& Hernquist~(\cite{Ka96}),
with a photo-ionizing flux similar to the one advocated by
Haardt \& Madau~(\cite{Ha96}).
Radiation from quasars reionizes the Universe at redshift $z\simeq 6$
in this scenario (for details, see e.g.~Dav\'e, Hernquist, Katz \&
Weinberg \cite{Da99}).
We have treated star formation in the framework of an effective multi-phase
model for the interstellar medium (ISM), as detailed by
Springel \& Hernquist~(\cite{SprHerMultiPhase}).
In this model, rapidly cooling gas of high overdensity is assumed to give
rise to the formation of cold clouds, embedded in an ambient hot medium.
The cloud material forms the reservoir of baryons available for star
formation.
Massive stars are taken to explode as supernovae on a short timescale,
and to release their energy as heat to the ambient medium of the ISM.
Supernovae are also assumed to evaporate cold clouds, thereby establishing
a tight self-regulation cycle for the star-forming ISM.

In addition, we have included a phenomenological model for galactic outflows
in our simulations, motivated by the large body of observational evidence for
the ubiquitous existence of such galactic winds
(e.g.~Martin~\cite{Mar99}).
Galactic-scale outflows are thought to play a crucial role in the enrichment
of the low-density intergalactic medium with heavy elements and for the global
regulation of star formation, and they may have influenced the properties of
intragroup and intracluster gas in a decisive way.
Note that without the inclusion of such strong feedback processes, simulations
tend to substantially overpredict the luminosity density of the universe
(Balogh et al.~\cite{Bal01}).
However, the model for galactic winds that we use here is able to reduce the
global efficiency of cooling and star formation sufficiently to match these
constraints (Springel \& Hernquist~\cite{SprHerSFR}).

In this model, each star-forming galaxy drives a wind with a mass-outflow rate
equal to two times its star formation rate, and with a wind-velocity of
$v_w=484\, {\rm km\,s^{-1}}$. These choices are deliberately `extreme' in the
sense that the total kinetic energy of the wind is of the order of the energy
released by the supernovae.
However, these parameters are quite typical for the observed properties of
outflows from star-forming disks.  Whether or not the wind can escape from a
galaxy depends primarily  on the depth of its dark matter potential well.
Halos with virial temperatures below $\sim 10^6\,$K can lose some  of their
baryons to an outflowing wind, while more massive halos will contain the wind,
such that the wind becomes progressively more unimportant for the regulation
of star formation in more massive objects.

\begin{figure}
\begin{center}
\resizebox{3.5in}{!}{\includegraphics{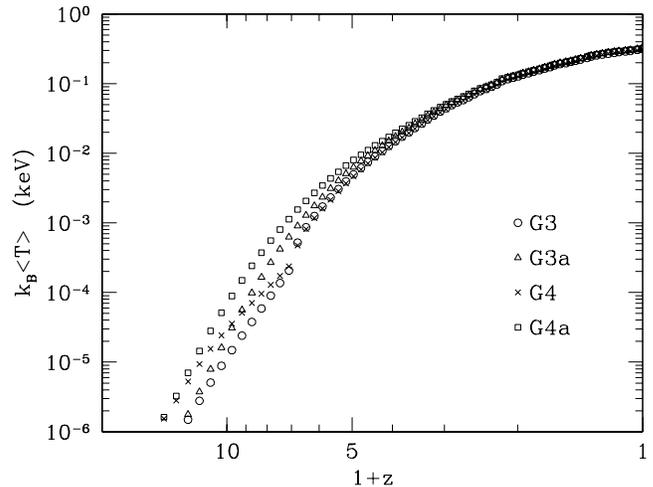}}
\end{center}
\caption{\footnotesize%
The evolution of the mean (mass weighted) temperature in the 4 simulations.
Parameters for the simulations can be found in Table~\protect\ref{tab:sims},
an `a' after the name indicates that feedback and cooling were not included.}
\label{fig:tz}
\end{figure}

The simulations considered here are part of a larger simulation program which
shows that the particular feedback model we employ allows a numerically
converged prediction for the total star formation density of the universe as a
function of epoch, provided all star-forming halos above virial temperatures
of $10^4\,$K can be resolved.
Of course, the simulations analyzed here cannot reach such a high mass
resolution, because they need to model a volume large enough to contain rich
clusters of galaxies.
It is thus inevitable that part of the star formation is lost due to
resolution limitations.  This can be a sizable effect.
Springel \& Hernquist~(\cite{SprHerSFR}) show that about 50\% of all stars
are born in halos with virial masses below $10^{12}\,h^{-1}{\rm M}_\odot$,
which corresponds roughly to the mass of the smallest halos in our G4
simulation with reliable estimates for the star formation rate.
Note that the `unresolved' star formation in small halos is also the one that
would predominantly be able to drive truly outflowing winds.
It should therefore be kept in mind that the effects of galactic outflows on
the intracluster medium are likely going to be underestimated by the present
simulations.

The simulations were performed on an Athlon-MP cluster at the Center
for Parallel Astrophysical Computing (CPAC) at the Harvard-Smithsonian
Center for Astrophysics. We used a modified version of the
{\small GADGET}-code (Springel, Yoshida \& White~\cite{SprYosWhi}), and
integrated the entropy as the independent thermodynamic variable
(Springel \& Hernquist~\cite{SprHerEntropy}).

\begin{table}
\begin{center}
\begin{tabular}{ccccc}
Model &      $N_p$      & $M_{DM}$ & $M_{\rm gas}$ & $\epsilon$ \\
G3    & $2\times 144^3$ & 2.5      & 0.4           &   18       \\
G4    & $2\times 216^3$ & 0.7      & 0.1           &   12       \\
\end{tabular}
\end{center}
\caption{The parameters of the simulations run.  The simulation box was
$100\,h^{-1}$Mpc on a side.  Particle masses are given in units of
$10^{10}\,h^{-1}\,M_\odot$.  Gravity was softened with a spline on a comoving
scale $\epsilon$ (in $h^{-1}{\rm kpc}$), with the force being fully Newtonian
beyond $2.8\,\epsilon$ and the potential of a single particle being
$-G m / \epsilon$ at zero lag, equivalent to a Plummer-softening on the same
scale.}
\label{tab:sims}
\end{table}

In all our runs, the full simulation box was output every $100\,h^{-1}$Mpc
along the line-of-sight between redshifts $z\simeq 19$ and $z=0$.
This corresponds to 77 dumps per model.  We made maps\footnote{The missing
factor of $h$ in our code which caused the maps from
Springel et al.~(\cite{SprWhiHer}) to be in error has been corrected here.}
in a manner similar to da~Silva et al.~(\cite{SBLT00}) as described in our
earlier paper (Springel et al.~\cite{SprWhiHer}).  To recap briefly we
stacked the various outputs from the boxes back along the line of sight,
with each box randomly translated and oriented in the direction of either
the $x$, $y$ or $z$ axis.
Then, a grid of $512^2$ rays subtending a constant angle of $1^{\rm o}$ from
the observer was traced through the boxes starting at $z=19$.

We produced maps of the $y$-parameter, the Doppler $b$-parameter,
the projected gas density and the thermal bremsstrahlung emission.
In the adiabatic simulations we do not track the ionization
fraction, $x_e$, dynamically.  Instead, we set it to 1.158, corresponding to
full ionization, for temperatures above $10^4\,$K and take the gas at lower
temperature to be neutral.
The extra physics models self-consistently track $x_e$ and this is used in
the computation.

It is also convenient to have a `cluster catalog' from the simulations.
We produced this by running a friends-of-friends group finder
(e.g.~Davis et al.~\cite{DEFW}) with a linking length of $b=0.15$
(in units of the mean inter-particle spacing).
The FoF algorithm partitions the particles into equivalence classes
by linking together all particle pairs separated by less than a distance $b$.
We define the center of each halo as the position of the potential minimum,
calculating the potential using only the particles in the FoF group.  This
proved to be more robust than using the center of mass, as the potential
minimum coincided closely with the density maximum for all but the most
disturbed clusters.  Additionally, the position of the center was very
insensitive to the presence or absence of the particles near the outskirts
of the halo, and thus to the precise linking length used.
For each of the halos we computed the mass, $M_{200}$, enclosed within a
radius, $r_{200}$, interior to which the density contrast was 200 times the
critical density.  We shall sometimes refer to this as the virial radius,
and denote its angular extent at the distance of the cluster by $\theta_{200}$.

\section{Results} \label{sec:results}

\begin{figure}
\begin{center}
\resizebox{3.5in}{!}{\includegraphics{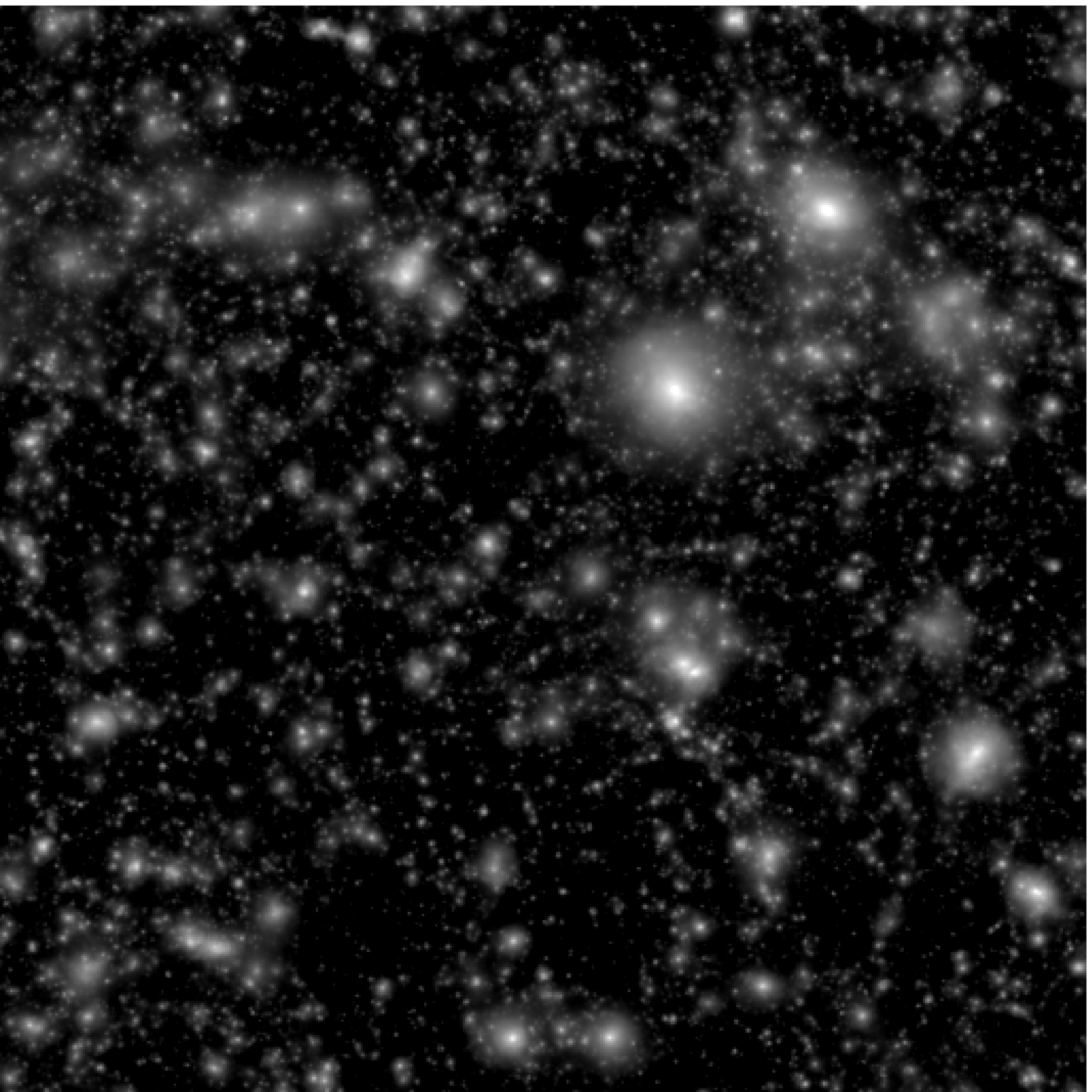}}
\resizebox{3.5in}{!}{\includegraphics{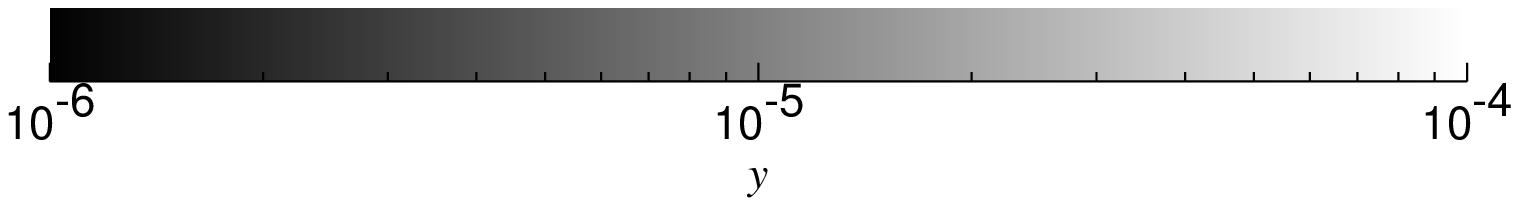}}
\end{center}
\caption{\footnotesize%
One of our Compton $y$ maps, from the simulation G4 including cooling and
feedback as described in the text.  The corresponding map from the adiabatic
simulation looks almost identical to the eye.}
\label{fig:maps}
\end{figure}

The evolution of the mean (mass weighted) temperature in the simulations is
shown in Fig.~\ref{fig:tz}.  The formation of structure in the universe leads
to a steady heating with the mean temperature today being approximately
$1/3$keV.  We can see the effects of cooling most clearly at early times in
the suppression of $\langle T\rangle$ in runs G3 and G4 with respect to G3a
and G4a, but by $z=0$ most of the differences have disappeared.
The slight jump in $\langle T\rangle$ at $z\sim 6$ is due to reionization by
quasars in this model, which raises the background radiation field and leads
to increased photo-heating of the gas in the universe.

We show a typical example of our thermal SZ maps in Fig.~\ref{fig:maps}.
These maps were produced by coadding a large number of partial maps,
each giving the contribution of one of the boxes that we stacked along
the photons' path.
We made 15 maps for each of the 4 simulations, using the same set of
random offsets and orientations for each of the 4 simulations.
As we noted before (Springel et al.~\cite{SprWhiHer}) filaments, while
prominent in the partial maps, are largely hidden in the high level of
background arising from the summation over many of these structures
(see also Croft et al.~\cite{Cro}).

The mean Comptonization we predict from our 15 maps is
$2.4\times 10^{-6}$ and $2.5\times 10^{-6}$ for the lower and higher
resolution adiabatic models respectively.
This becomes $2.2\times 10^{-6}$ and $2.1\times 10^{-6}$ for the runs
with cooling and feedback.
Thus we see that the effect of cooling is more important than heating in
aggregate for the gas contributing to the mean $y$ distortion and that
our models easily pass the constraint from FIRAS (Fixsen et al.~\cite{FIRAS}).

\subsection{Power spectra}

\begin{figure}
\begin{center}
\resizebox{3.0in}{!}{\includegraphics{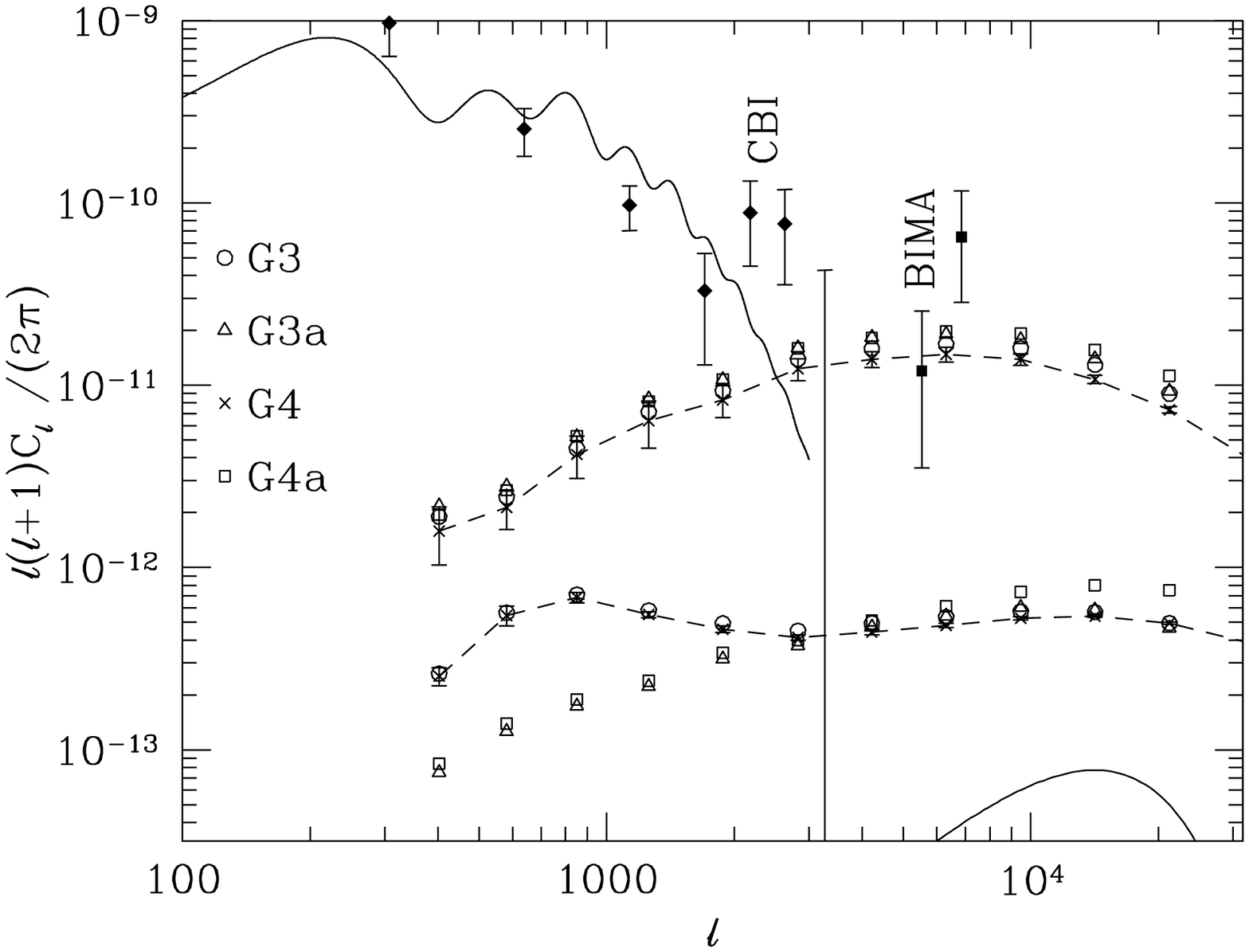}}
\resizebox{3.0in}{!}{\includegraphics{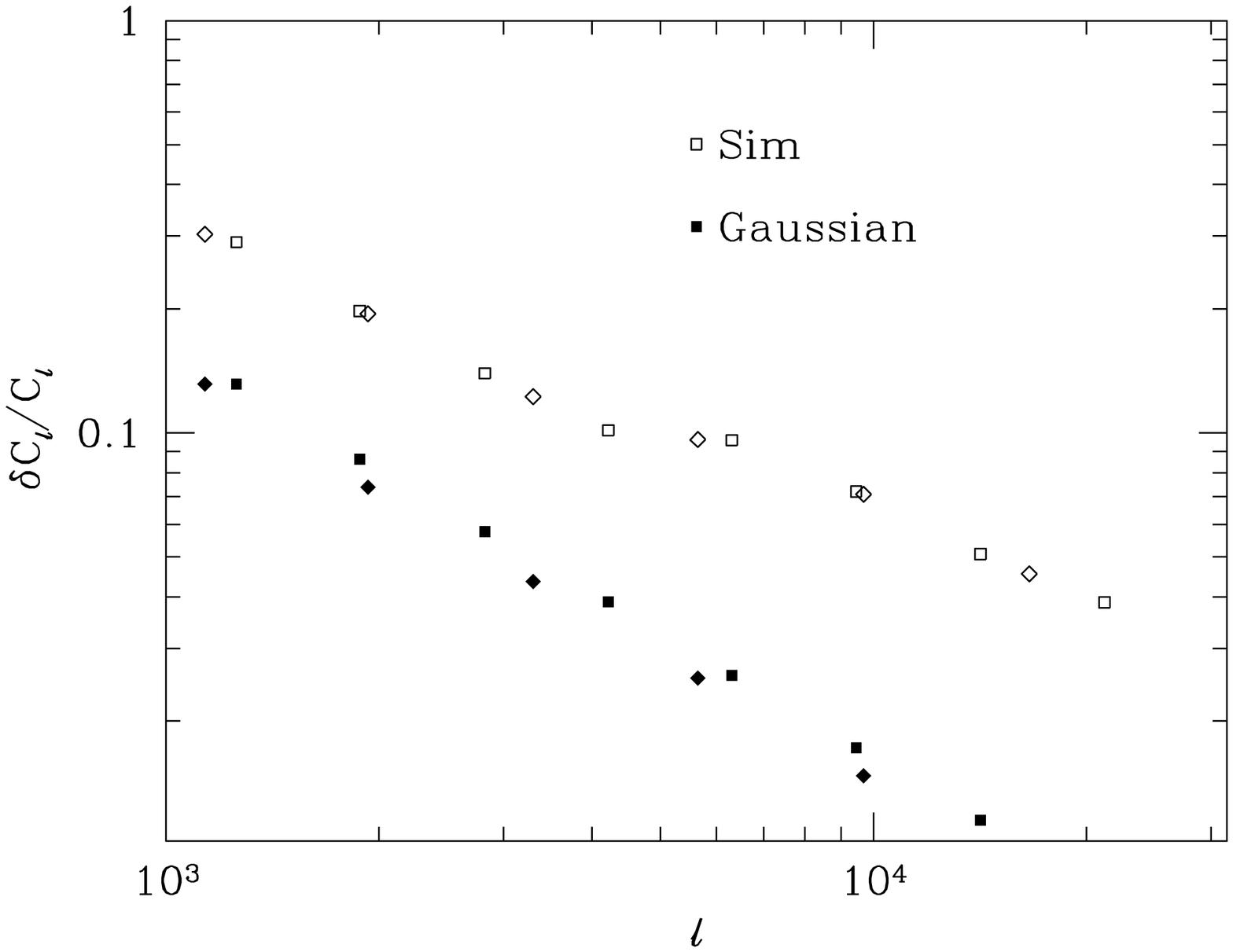}}
\resizebox{3.0in}{!}{\includegraphics{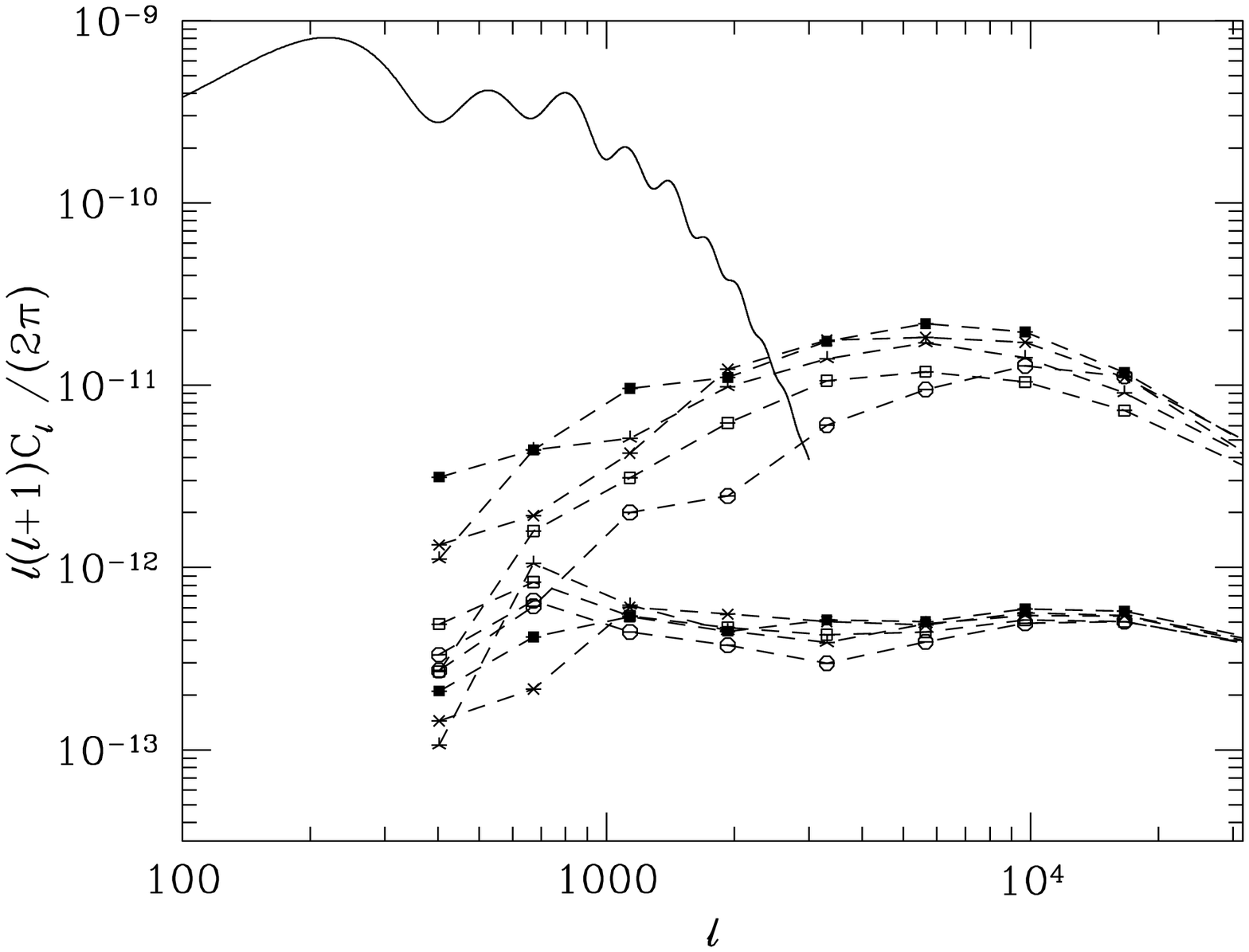}}
\end{center}
\caption{\footnotesize%
(Top)
The angular power spectra for the thermal and kinetic SZ effects
averaged over 15 maps for each of our 4 simulations.  Parameters for the
simulations can be found in Table~\protect\ref{tab:sims}, an `a' after the
name indicates that feedback and cooling were not included.
The upper solid curve shows the primary anisotropy signal for this model
(including the effects of lensing).  The symbols the mean SZ signal and
the error bars the standard deviation of the power from the 15 maps.
A dashed line connects the results of simulation G4.
The lower solid curve is an estimate of the effects of patchy reionization
assumed to occur at $z=6$.
The points with error bars are observational results from BIMA and CBI
(see text).
(Middle) The fractional error in the thermal effect, per binned $C_\ell$,
obtained from the variance of the 15 simulated fields (open symbols) compared
to the expectations for a Gaussian sky (filled symbols).
Symbol types denote two different binnings of the power spectrum.
The non-Gaussian nature of the SZ signal increases the field-to-field
scatter over the Gaussian expectation.
(Bottom) The angular power spectra for 5 of our fields, showing that the
field-to-field fluctuation is primarily in the amplitude, and not the
shape, of the spectra.}
\label{fig:lcl}
\end{figure}

For each map we computed a number of statistics, and we in general
averaged the results for 15 random lines of sight to reduce
field-to-field variance.  Here we present the angular power spectrum of
the thermal and kinetic SZ effects (Fig.~\ref{fig:lcl}).

A comparison of the low and high resolution simulations indicates that for
the adiabatic case we have converged in the angular power spectrum out to
$\ell\sim 20,000$.  These results are also in excellent agreement with our
earlier estimates (Springel, White \& Hernquist~\cite{SprWhiHer}) once we take
into account the slight differences in the cosmological model.
For the runs including cooling we are more sensitive to the effects of
finite mass and spatial resolution.  The higher resolution simulation shows
a lower power spectrum than the lower resolution spectrum, indicating that
cooling has been able to proceed more efficiently when the simulation can
correctly resolve denser structures.
However, the final results are quite insensitive to extra physics included
in these models, as might have been expected if the bulk of the signal is
coming from gas outside the core of the cluster, and the gas content of the
intracluster medium is not significantly depleted by gas cooling.
In this respect our results differ from those of
da Silva et al.~(\cite{SKLTPB})
who found a significant effect on the SZ signal if they include cooling in
their simulations.
The addition of star formation and energy injection in our simulations
stabilizes the runaway cooling which otherwise occurs and significantly
reduces the impact of the extra physics on the SZ effect.
In the absence of such feedback the cooling is only mitigated by numerical
resolution, leading to an excess of cold gas and stars in conflict with
observations.

The power spectra we predict are in accord with recent measurements of
arcminute angular scale anisotropy in the CMB at BIMA
(Dawson et al.~\cite{BIMA1}; \cite{BIMA2}) though lower than the detection
quoted by CBI (Mason et al.~\cite{CBI1}; Bond et al.~\cite{Bond}) from their
deep field
(the data from mosaic observations, Pearson et al.~\cite{CBI2}, do not
significantly constrain the signal at high-$\ell$).
To bring the adiabatic model into agreement with the CBI points would require
us to e.g.~increase $\sigma_8$ to 1--1.2 (or dramatically increase $\Omega_B$;
see e.g.~Seljak, Burwell \& Pen~\cite{SelBurPen} or Bond et al.~\cite{Bond}).
The inclusion of cooling and feedback lowers our predictions by $\sim 30\%$ on
CBI scales, which could be compensated for by an additional $\sim\,4\%$ change
in $\sigma_8$.

We have also estimated the amount by which the evident non-Gaussian nature of
the SZ signal increases the field-to-field variance in the power spectrum.
For a Gaussian signal the fractional error per mode in the power spectrum
is $\sqrt{2}$, so we expect the standard deviation in the binned power to
be $\sqrt{2/N}\,C_\ell$ where $N$ is the number of modes in the bin.
If we compare this expectation to the standard deviation computed from the 15
maps (Fig.~\ref{fig:lcl}), we find that the latter is higher by a factor of 2
(at $\ell\sim 10^3$) to 5 (at $\ell\sim 10^4$) due to the contributions from
the 3- and 4-point functions (e.g.~Meiksin \& White~\cite{MeiWhi}, \S2).
Since the non-Gaussianity leads to mode-coupling, the exact ratio depends
slightly on the binning chosen -- we show two choices.
We caution the reader that all of the maps are created from the same underlying
simulation volume, and so are not fully independent.  This makes the estimates
at lower $\ell$ somewhat uncertain.
We do not have enough maps to accurately calculate the off-diagonal elements
of the correlation matrix (which converge slowly compared to the diagonal
elements), however the correlations are strong.
As for the non-linear mass power spectrum, the primary fluctuation from field
to field is in the amplitude, and not the shape, of the power spectrum.  We
show an example from 5 of our fields in Fig.~\ref{fig:lcl}.

Finally it is of interest to ask where the majority of the thermal SZ signal
is coming from in the angular power spectrum -- i.e.~diffuse gas or gas within
halos.
To answer this we remade two of our maps excluding all gas with density
smaller than $100\,\bar{\rho}_{\rm gas}$.  Gas in halos should have higher
density than this.
The resulting maps are indistinguishable by eye, even `blinking' between them.
There is a small (25\%) suppression of power for $\ell<2000$ in the maps
with the density cut, but the suppression is $<2\%$ by $\ell=6000$ where the
signal peaks.
This suggests (da Silva et al.~\cite{SKLTPB})
that even the large-angle signal is coming primarily from relatively high
density gas along the line-of-sight --- the virialized halos -- and not from
the diffuse gas or filamentary structures (the `cosmic web').
If we increase the cut even further, to $10^3\bar{\rho}_{\rm gas}$, we start
to significantly reduce the thermal SZ signal up to $\ell\sim 10^4$,
indicating that the dense cores of the halos do not dominate the thermal
SZ signal.

\subsection{Source counts}

A glance at Fig.~\ref{fig:maps} indicates that the SZ effect is dominated by
discrete sources and that the pixel distribution is highly non-Gaussian.
Thus a power spectral analysis contains only a small part of the information
in the map.  To obtain an understanding of the SZ source sizes and strengths
we have used the source detection software SExtractor
(Bertin \& Arnouts \cite{Bertin96}) to count individual sources, and to
measure their strength.
We used default settings for the detection algorithm of the image processing
software and let it estimate background and `noise' levels automatically.
Note that our maps are in principle noise-free, so what is interpreted as
noise in this procedure effectively arises from limitations due to source
confusion, which leads to non-detections of some of the faintest and smallest
sources.
In a typical analysis with SExtractor, we resolve slightly more than half of
the total SZ signal into discrete sources.

\begin{figure}
\begin{center}
\resizebox{3.5in}{!}{\includegraphics{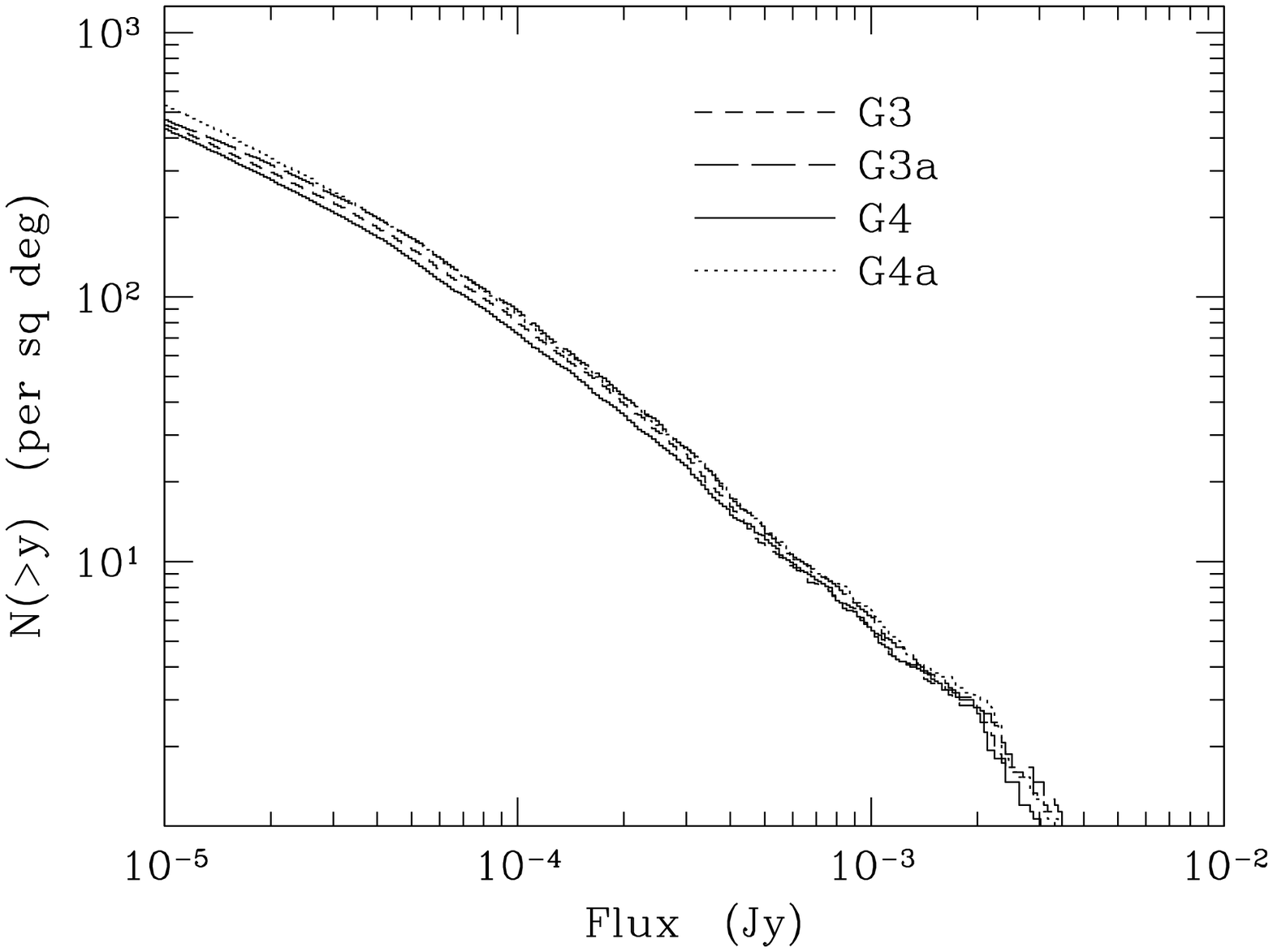}}
\resizebox{3.5in}{!}{\includegraphics{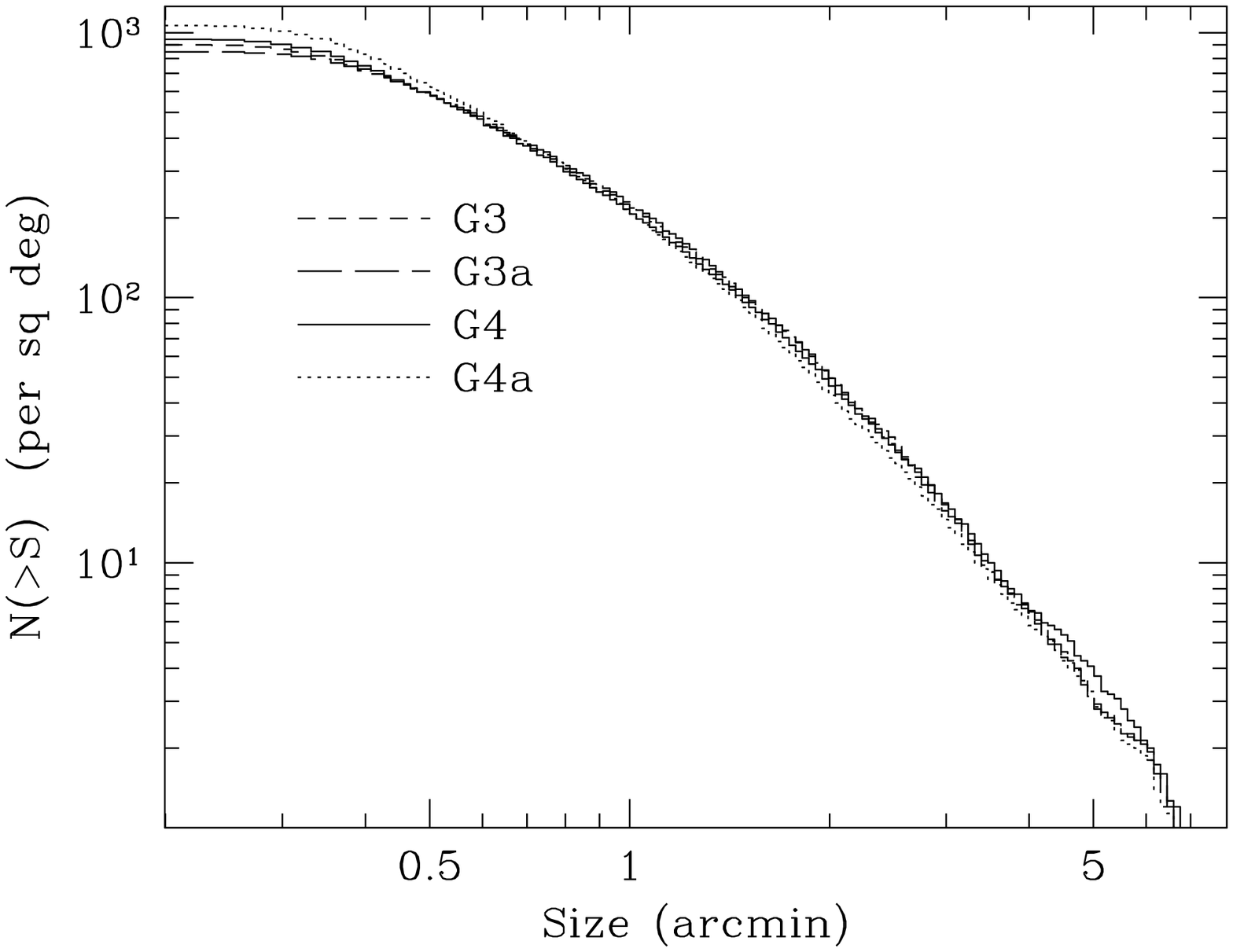}}
\end{center}
\caption{\footnotesize%
Upper panel: the cumulative distribution of source fluxes at $\nu=30$GHz.
Lower panel: the cumulative distribution of source sizes.}
\label{fig:sources}
\end{figure}

We measure the strength of a source as the monochromatic brightness change
\begin{equation}
  S_\nu = \int_\Omega \Delta B_\nu \, {\rm d}\Omega
        = f(x) B_{\nu} \int_\Omega y(\theta) \, {\rm d}\Omega
\end{equation}
of the CMB integrated over the solid angle of the source. Here $B_\nu$ is the
Planck spectrum of the primary CMB, and $f(x)$ is the spectral function 
\begin{equation}
  f(x)= x \frac{{\rm e}^x}{{\rm e}^x-1}
    \left( x{{\rm e}^x+1\over {\rm e}^x-1}-4 \right),
\end{equation}
with $x=h\nu/kT_{\rm CMB}$ and we have taken $\nu=30$GHz.

In Fig.~\ref{fig:sources}, we show the cumulative source counts per square
degree found in the simulations.
Again we find remarkably good agreement between all of the simulations, with
the cooling runs predicting fewer sources at the bright end by 20-30\% and
at the faint end by 30-50\% (at fixed flux).
This is not terribly surprising at the bright end where the source counts
are more massive clusters.  In this case the bulk of the ICM is at too low
a density to cool efficiently.
However our calculations suggest that this holds true even for the relatively
weak SZ sources.
Cooling and feedback thus appear to be a small effect on the clustering and
number counts of sources, with the gas dominating the SZ signal being more
prone to cooling than heating in our model.

\begin{figure}
\begin{center}
\resizebox{3.5in}{!}{\includegraphics{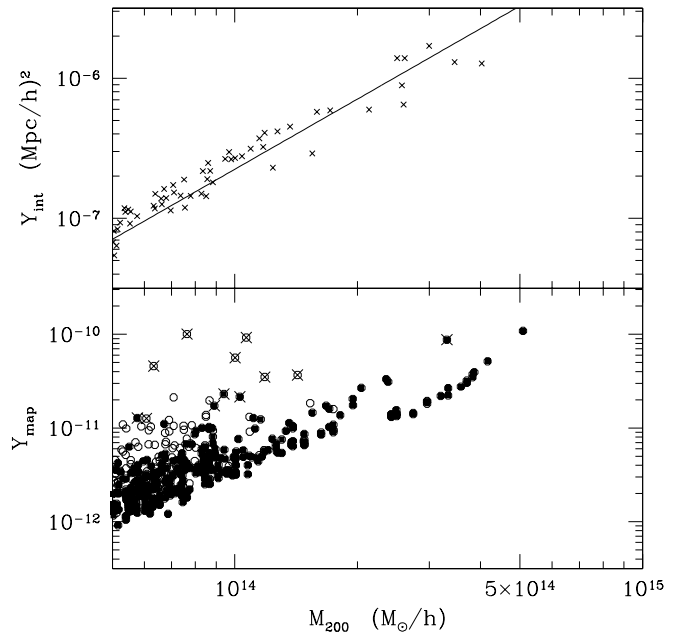}}
\end{center}
\caption{\footnotesize%
The relation between the total decrement, $Y$, and cluster mass $M_{200}$.
The upper panel shows the decrement calculated {}from the material within
half the virial radius $r_{200}$ for each cluster in the simulation volume at
$z=0$.  The straight line indicates $Y_{\rm int}\propto M_{200}^{5/3}$.
The lower panel indicates the decrement obtained directly from the
noise free maps by summing pixels centered on the known locations of clusters
(excluding clusters near the edge of the map).
Open circles indicate clusters whose disk overlapped that of a more massive
cluster in the map.  Filled circles are `isolated' clusters.  Crosses indicate
clusters within $600\,h^{-1}$Mpc of the observer.  The source of the scatter
is discussed in the text.}
\label{fig:ym}
\end{figure}

Each of these bright sources is associated with at least one massive halo in
the simulation.  It is instructive to pursue this association further.
In Fig.~\ref{fig:ym} we show a scatter-plot of source strength vs.~halo mass,
both intrinsic and from the maps.
For the `intrinsic' effect we calculated the SZ signal coming from the most
massive clusters in our box at $z=0$.
Specifically we summed the SZ signal within half of $r_{200}$ to obtain the
total decrement intrinsic to the cluster:
\begin{equation}
  Y_{\rm int} = (\gamma-1)(1-Y_p) {\sigma_T}
     \sum_j \mu_j x_j {m_j\over m_e} {u_j\over c^2}
\label{eqn:Ydef}
\end{equation}
where $\gamma=5/3$, $Y_p=24\%$, $\sigma_T$ is the Thomson cross section,
$m_e$ the electron mass, $c$ the speed of light,
$\mu_j$ is the mean molecular weight of particle $j$, $x_j$ is the fractional
ionization (relative to hydrogen), $m_j$ is the mass and $u_j$ the internal
energy.
The sum is over all particles within\footnote{The degree of correlation is
almost independent of the exact aperture used.} $r_{200}/2$ of the potential
minimum in the group, which is the region dominating the SZ emission.
Since $y$ is dimensionless we quote $Y_{\rm int}$ in units of
$(h^{-1}{\rm Mpc})^2$.
As the total signal scales as $M\times T$ and for self-similar clusters
$T\sim M^{2/3}$ we expect $Y_{\rm int}\sim M^{5/3}$ which is a good fit to
our simulations.
The upper panel of Fig.~\ref{fig:ym} shows that there is a very tight relation
between $Y_{\rm int}$ and $M_{200}$, as has been noted before by
e.g.~Holder et al.~(\cite{HMCEL}).
It is this relation which is behind the claims that a $y$-selected sample will
be close to mass limited.

However, $Y_{\rm int}$ is not an observable.
To mimic something which is closer to what an observation could probe we
first found all of the clusters in the simulation which fell in the
line-of-sight of a particular map.
For each of these we integrated the Compton $y$ parameter values in pixels
which lay within half $\theta_{200}$ of the cluster center to obtain
$Y_{\rm map}=\int d\Omega\, y$.
As the lower panel of Fig.~\ref{fig:ym} shows, there is a large amount of
scatter in this $Y_{\rm map}-M$ relation, even for our noise free, high
resolution maps.
This scatter is approximately constant as we vary the aperture used to define
$Y_{\rm map}$ and can be traced to 3 prime sources.
First, there is evolution in the $M-T$ relation with redshift, and the
clusters we have plotted are from a range of redshifts.
Second, clusters are not spherical, leading to differences in signal when
viewed from different orientations.
Finally, the signal we see is the projection of SZ signal from the entire
line-of-sight\footnote{This contribution was not included in the work of
Holder et al.~(\cite{HMCEL}) who simulated only gas within the virial regions
of their sources.}.
Since the signal is not dimmed by distance, objects all along the line-of-sight
can contribute a significant amount to the signal in any pixel.
A similar `confusion' effect operates for optical and weak lensing cluster
searches
(White \& Kochanek~\cite{WhiKoc}; White, van Waerbeke \& Mackey~\cite{WvWM}).
Since clusters form in high density regions, with a lot of dense material
nearby, there is a non-negligible chance that some lines of sight will be
enhanced by non-cluster material.
The extreme outliers in Fig.~\ref{fig:ym} are all clusters closer to the
observer than about $600\,h^{-1}$Mpc, whose virial radii subtend a large
angle and are most likely to suffer from this projection effect.  These
clusters are marked with crosses in the figure.

To isolate this effect further we have computed the integrated SZ signal
coming from spheres of gas, centered on rich clusters, viewed from many
random directions.  Specifically, we first extract all of the gas particles
in the simulation within a distance $r_{\rm cut}$ of the minimum of the
cluster potential.  From this set we compute, using Eq.~(\ref{eqn:Ydef}),
the total $Y$ for all particles whose 2D projected radius $R<r_{200}/2$
for several lines-of-sight.  We also compute $Y$ from the subset of the
particles contained within our FoF group.

\begin{figure}
\begin{center}
\resizebox{3.5in}{!}{\includegraphics{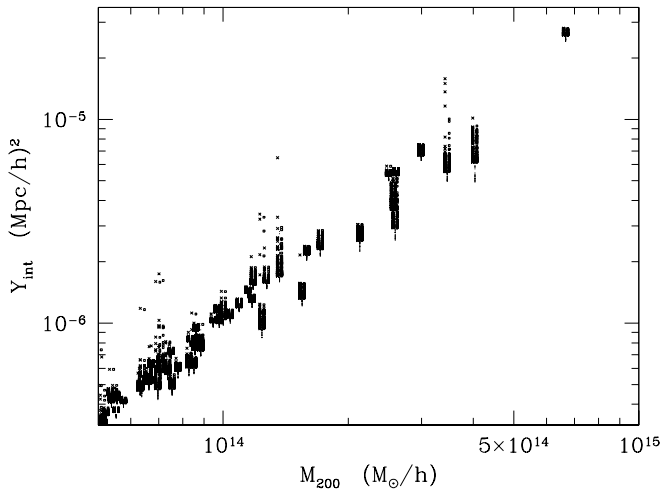}}
\end{center}
\caption{\footnotesize%
The scatter in the $Y_{\rm int}-M$ relation coming from material close to
the clusters.  $Y_{\rm int}$ is calculated for each cluster in the simulation
volume at $z=0$ when viewed down 64 random lines of sight.  Only material
within $R_{200}/2$ in projection was included.
Crosses include all material with $r_{\rm cut}=10\,h^{-1}$Mpc, open squares
$r_{\rm cut}=5\,h^{-1}$Mpc and dots only the material belonging to the FoF
group.  Points have been offset (horizontally) for clarity.}
\label{fig:ynear}
\end{figure}

Our results are shown in Fig.~\ref{fig:ynear} for $r_{\rm cut}=5\,h^{-1}$Mpc
and $r_{\rm cut}=10\,h^{-1}$Mpc.
For several clusters there is a large scatter between lines-of-sight depending
on the presence or absence of material near the cluster.  The scatter is
larger for the larger radius, indicating that it is not just neighboring
material which drives the scatter.
If we include only those particles within the FoF group we still see a large
scatter, though the SZ signal is noticeably depressed.
Thus, some of the scatter is coming from asphericity of the cluster, but a
non-negligible fraction of the signal is coming from non-cluster material
along some lines-of-sight.

The large scatter that we see in the SZ signal at fixed mass has implications
for SZ searches for galaxy clusters -- any method which uses primarily SZ flux
in target selection will not be mass limited.  In fact, the steeply falling
mass functions predicted by hierarchical theories would imply that some
high-flux candidates are likely to be low mass groups with abnormally high
SZ signal rather than the intrinsically rarer rich clusters.
If this is true, a blind SZ search should turn up many groups in addition to
the desired cluster sample.

\section{Conclusions} \label{sec:conclusions}

We have performed a sequence of high resolution hydrodynamic simulations
of structure formation in a $\Lambda$CDM model to investigate the effect
of extra physics on the thermal and kinetic Sunyaev-Zel'dovich effects.
Including only adiabatic gas physics our simulations of the thermal effect
are converged down to sub-arcminute scales and agree well with our earlier
work.  For the runs with extra physics we see an increased efficiency of
cooling in the higher resolution simulation which leads to a lowering of
the mean Compton $y$ parameter and a slight depression in the SZ angular
power spectrum.
Overall however the effects of including the extra physics on the 1- and
2-point functions are slight, indicating that the SZ effect is not overly
sensitive to the (uncertain) thermal history of the gas.

The maps are noticeably non-Gaussian and dominated by discrete sources.
This non-Gaussianity manifests itself in increased scatter, by a factor
of 2-6, in $C_\ell$ from field-to-field compared to the Gaussian expectation.

The properties of the sources in the map are quite insensitive to the extra
physics we have included also, and we are thus able to make robust statements
about their detection.
The brightest SZ sources have a density of $\sim\,$1 per square degree and
their detection requires an angular resolution of $\sim\,$1$'$, and the
capability to separate brightness fluctuations of size $\sim\,$10$\,{\rm mJy}$
{}from the primary CMB fluctuations.

We have found that there is significant scatter in the SZ fluxes within
apertures centered on clusters.  This scatter has 3 sources: evolution of
the mass temperature relation, asphericity in the matter distribution and
line-of-sight projection.  The existence of this scatter means that surveys
using only SZ flux will select objects with a wide range of masses, including
lower mass groups.

\bigskip
\acknowledgments
We would like to acknowledge useful conversations with Kyle Dawson,
Bill Holzapfel, Adrian Lee and Saul Perlmutter.
This work was supported by
ACI 96-19019, AST 98-02568, AST 99-00877, AST 00-71019.
M.W. was supported by a NASA Astrophysical Theory Grant and a
Sloan Fellowship.
The simulations were performed at the Center for Parallel Astrophysical
Computing at the Harvard-Smithsonian Center for Astrophysics.

\end{document}